\documentclass[sigconf]{acmart}

\usepackage[utf8]{inputenc}
\usepackage{bm}
\usepackage{float} 

\usepackage{graphicx}
\usepackage{subfigure}

\usepackage{verbatim}

\floatname{algorithm}{Algorithm}
\usepackage{algorithm}
\usepackage{algpseudocode}

\usepackage{booktabs} 
\usepackage{xfp}
\usepackage{mathtools}
\usepackage{bbm}
\usepackage{amsmath}
\usepackage{gensymb}  

\usepackage{microtype}

\AtBeginDocument{%
  \providecommand\BibTeX{{%
    \normalfont B\kern-0.5em{\scshape i\kern-0.25em b}\kern-0.8em\TeX}}}

\copyrightyear{2023} 
\acmYear{2023} 
\setcopyright{rightsretained} 
\acmConference[UbiComp/ISWC '23 Adjunct ]{Adjunct Proceedings of the 2023 ACM International Joint Conference on Pervasive and Ubiquitous Computing \& the 2023 ACM International Symposium on Wearable Computing}{October 8--12, 2023}{Cancun, Quintana Roo, Mexico}
\acmBooktitle{Adjunct Proceedings of the 2023 ACM International Joint Conference on Pervasive and Ubiquitous Computing \& the 2023 ACM International Symposium on Wearable Computing (UbiComp/ISWC '23 Adjunct ), October 8--12, 2023, Cancun, Quintana Roo, Mexico}\acmDOI{10.1145/3594739.3612912}
\acmISBN{979-8-4007-0200-6/23/10}

\begin{document}

\title[Cooperative Multi-Type MARL for Resource Management in SAGIN]{Cooperative Multi-Type Multi-Agent Deep Reinforcement Learning for Resource Management in Space-Air-Ground Integrated Networks}


\author{Hengxi Zhang}
\email{zhanghx20@mails.tsinghua.edu.cn}
\orcid{1234-5678-9012}
\affiliation{%
  \institution{Tsinghua-Berkeley Shenzhen Institute}
  \country{Shenzhen, Guangdong, China}
}

\author{Huaze Tang}
\email{thz21@mails.tsinghua.edu.cn}
\affiliation{%
  \institution{Tsinghua-Berkeley Shenzhen Institute}
  \country{Shenzhen, Guangdong, China}
}

\author{Wenbo Ding}
\authornote{Corresponding author.}
\email{ding.wenbo@-sz.tsinghua.edu.cn}
\affiliation{%
  \institution{Tsinghua-Berkeley Shenzhen Institute}
  \institution{RISC-V International Open Source Laboratory}
  \institution{Shenzhen, Guangdong, China}
  \institution{Shanghai Artificial Intelligence Laboratory}
  \country{Shanghai, China}
}

\author{Xiao-Ping Zhang}
\email{xiaoping.zhang@ee.ryerson.ca}
\affiliation{%
  \institution{Tsinghua-Berkeley Shenzhen Institute}
  \institution{RISC-V International Open Source Laboratory}
  \institution{Shenzhen, Guangdong, China}
  \institution{Department of Electrical, Computer and Biomedical Engineering, Ryerson University}
  \country{Toronto, Canada}
}


	\begin{abstract}
        The Space-Air-Ground Integrated Network (SAGIN), integrating heterogeneous devices including low earth orbit (LEO) satellites, unmanned aerial vehicles (UAVs), and ground users (GUs), holds significant promise for the advancing smart city applications.
        However, resource management of the SAGIN is a challenge requiring urgent study in that inappropriate resource management will cause poor data transmission, and hence affect the services in smart cities.
        In this paper, we develop a comprehensive SAGIN system that encompasses five distinct communication links and propose an efficient cooperative multi-type multi-agent deep reinforcement learning (CMT-MARL) method to address the resource management issue.
        The experimental results highlight the efficacy of proposed CMT-MARL, as evidenced by key performance indicators such as the overall transmission rate and transmission success rate. These results underscore the potential value and feasibility of future implementation of the SAGIN.
	\end{abstract}





\maketitle

\vspace{-.2cm}
\section{Introduction}
    The advent of 5G technology has ushered in a new era where the Internet of Things (IoT) serves as the backbone of numerous applications and services, such as intelligent transportation systems, home automation, and smart factories~\cite{IoT,shang2021computing, andrews2014will}. However, the burgeoning computational demands of IoT devices stretch the limits of existing wireless communication networks. Current terrestrial communication paradigms are ill-equipped to cope with this demand. Enter the Space-Air-Ground Integrated Network (SAGIN), a proposed heterogeneous system that amalgamates satellites, unmanned aerial vehicle (UAV)-based air systems, and terrestrial communication systems like base stations~\cite{liu2018space}. Its objective is to provide versatile network coverage and services.
    
    SAGIN has emerged as a potential solution to meet the mounting computational requirements of IoT services and applications \cite{hubenko2006improving, albuquerque2007global, pultarova2015telecommunications}. By integrating terrestrial systems with satellites and unmanned aerial vehicle (UAV)-based air systems, SAGIN offers a flexible and scalable approach that could potentially rise to these challenges \cite{xiong2016kind}.
    However, the implementation of SAGIN within IoT services is not without its hurdles \cite{el2016resource}. The primary obstacle lies in the inherent high mobility of the air system. This mobility results in dynamic and often unpredictable channel conditions and coverage areas. Moreover, the ground connections in current approaches are frequently overlooked. Exceptions are made only for connections between ground users and UAVs, an approach that can lead to inefficiencies and limit the full utilization of the integrated network~\cite{kato2019optimizing}. 
    An additional complication arises from the heterogeneity of the SAGIN subsystems. Each subsystem — terrestrial, aerial, and space — uses a unique communication interface, reflecting its particular technological requirements and constraints~\cite{liu2018space}. Furthermore, channels between these various subsystems possess distinct properties, further complicating communication and integration. 
    Therefore, it is imperative to develop a discerning resource management policy for SAGIN. 

    The rapid evolution of machine learning techniques offers potential solutions to these challenges~\cite{zhou2020deep}. Among these, reinforcement learning (RL) stands out as a reward-centric approach designed to tackle combinatorial optimization problems. An RL agent interacts with and learns from its environment to optimize toward an end-goal, guided by a predefined reward system~\cite{sutton2018reinforcement}. In some contexts, there may be multiple agents, or centralized learning methods may prove ineffective due to inherent environmental characteristics~\cite{zhang2021multi}. In these situations, the notion of multi-agent RL (MARL) emerges as a decentralized learning method. It has been increasingly adopted by researchers and institutions over recent years~\cite{foerster2016learning}. Additionally, there may exist more than one type of agent in the multi-agent system, where different types of agents need to utilize diverse behaviors to coordinate with others~\cite{subramanian2020multi}.
    In this work, we develop an ambiguous stochastic optimization (ASO) SAGIN communication model aimed at handling the system's overall resource management problem. We then employ a tailor-made MARL technique, dubbed Cooperative Multi-Type MARL (CMT-MARL), to elaborately study how the multi-agent system will perform in this dynamic SAGIN.
    \section{Heterogeneous System of Wireless Communication}
\begin{figure}[t] 
        \centering
        \includegraphics[width=.85\linewidth]{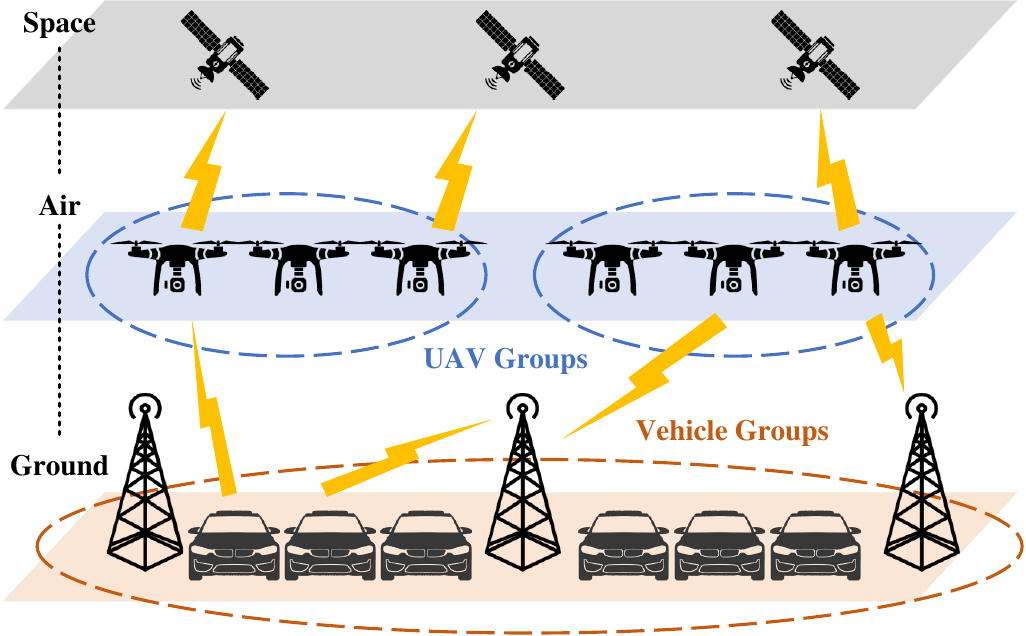}
        \caption{A schematic illustration of SAGIN.}
        \label{SAGI}
        \vspace{-.4cm}
    \end{figure}

	In this SAGIN system presented in Fig.~\ref{SAGI}, we consider four communication categories at three different altitudes, with a bunch of GUs (i.e. vehicles) and a BS on the ground layer, the UAV swarm hovering in the air layer and an LEO satellite in the space layer, and five classes of communication links among these categories. 
	The links started from GUs to BS, UAVs, and satellites are modeled as GU-to-BS (G2B) link, GU-to-UAV (G2U) link, and GU-to-Satellite (G2S) link, respectively.
	While each UAV can also establish two kinds of links from itself to BS and satellite, expressed similarly as the UAV-to-BS (U2B) link and UAV-to-Satellite (U2S) link.
	
	Both GU and UAV groups continuously generate a series of data packets that need to be transmitted to the remote servers eventually.
	The BS and satellite are considered as the terminal transmission devices for each GU and UAV to transmit the data packets in the SAGIN, where the UAV in the air layer can be utilized as an intermediary for each GU in the ground to transmit the packet when the G2S or G2B link is temporally crowded or considering the transmit speed of them are relatively low.
	Specifically, GUs and UAV swarm involve $N_1(=|\mathcal{N}_1|)$ and $N_2(=|\mathcal{N}_2|)$ homogeneous individuals respectively, and can be further treated as heterogeneous system $\mathcal{N}(=\mathcal{N}_1\cup\mathcal{N}_2)$, where $\mathcal{N}_1=\{GU_1, GU_2,\dots, GU_{N_1}\}$ and $\mathcal{N}_2=\{UAV_1, UAV_2,\dots, UAV_{N_2}\}$.
	Each GU can select only one link from G2B, G2U, and G2S links to transmit data, while U2B or U2S are two links for each UAV to choose from. 
	Note that the G2U link can only be generated if the UAV selected by the GU agrees to make this connection.
    \subsection{Wireless Communication Links}
    \label{subsec:links}
    The unicast protocol is employed in this work, where each GU or UAV can only select one objective to transmit its message, and the transmission between the same category, such as GU-to-GU and UAV-to-UAV, is prohibited to prevent message congestion, following classic configuration of SAGIN \cite{zhang2017software}. 
    Since there exist three layers in SAGIN, we allocate two bandwidths $W_1$ and $W_2$ that support the low-altitude and high-altitude transmission, respectively, in which low-altitude bandwidth $W_1$ supports G2B, G2U, and U2B links while G2S and U2S links are on the high-altitude bandwidth $W_2$.

    \subsubsection{Low-Altitude Links}
    The low-altitude links, also known as the terrestrial links, are operated terrestrially, including G2B links, G2U links, and U2B links \cite{chandrasekharan2016designing, agiwal2016next}. The low-altitude links feature relatively lower bandwidth and frequency than high-altitude links and provide communication with lower latency and enhanced reliability due to low pathloss \cite{zhang2017software}. Taking into account the interfering channels within the sub-band, as discussed in \cite{liang2019spectrum}, the signal-to-interference-plus-noise ratio (SINR) for low-altitude links operating on the $h$-th sub-band (where $h\in[H]$) can be formulated as follows,
    \vspace{-5pt}
    \begin{equation}
        \gamma_i^{LOW}[h] = \frac{p_i^{LOW}[h] g_{i,d_i}[h]}{\sigma^2+ \phi_i^{LOW}[h]},
    \end{equation}
    where $p_i^{LOW}[h]~$ and $g_{i,d}[h]$ refer to the transmit power and the interfering channel from $i$-th $(i\in\mathcal{N})$ device (GU or UAV) to its designated communication endpoint denoted as $d_i$ (base station or another device) over $h$-th sub-band. $\phi_i^{LOW}[h]$ denotes the interference of low-altitude links and can be expressed as,
    $
            \phi_i^{LOW}[h]= \sum_{i^\prime\in\mathcal{N}_i}  x_{({i^\prime},d_{i^\prime})}[h] p_{i'}^{LOW}[h] g_{{i^\prime},d_{i^\prime}}[h],
    $
    where $p_{i^\prime}^{LOW}[h]$ and $g_{{i^\prime},d_{i^\prime}}[h]$ similarly represent the transmit power and the interfering channel  from $i^\prime$-th device to the communication destination $d_{i^\prime}$.
    $x_{({i^\prime},d_{i^\prime})}(\cdot)$ is the indicator function equal to 1 if the $h$-th sub-band is occupied and 0 otherwise.
    
    \subsubsection{High-Altitude Links}
    The high-altitude links, also known as the non-terrestrial links, encompass the connections between terrestrial devices \cite{casoni2015integration}. Specifically, the high-altitude links consist of G2U links and U2S links. These links enable communication and data exchange between the terrestrial devices (GUs and UAVs) and the satellite. Since the satellite can be utilized to support long-range transmission due to its wide horizon, the high-altitude link in the proposed SAGIN model can be treated as a relatively smaller BS with a relatively wide bandwidth for receiving a series of messages from both GUs and UAVs \cite{farserotu2000survey}.
    Therefore, we express the SINR of high-altitude link over $h$-th ($h\in[H]$) sub-band as,
    \vspace{-2pt}
    \begin{equation}
        \gamma_i^{HIGH}[h] = \frac{p_i^{HIGH}[h] g_{i,S}[h]}{\sigma^2+ \phi_i^{HIGH}[h]},
    \end{equation}
    \vspace{-2pt}
     where $p_i^{HIGH}[h]$ and $g_{i,S}[h]$, similar to the high-altitude link, are the transmit power and the interfering channel from $i$-th terrestrial device to the satellite over $h$-th sub-band.
     Considering there are only G2S and U2S links on the high-altitude bandwidth $W_2$, we use 
    $
            \phi_i^{HIGH}[h] = \sum_{i'\in\mathcal{N}^i}  x_{{i'},S}[h] p_{i'}^{HIGH}[h] g_{{i'},S}[h]
            $
    to indicate the interference power of high-altitude link from $i$-th device to the satellite over $h$-th sub-band.

    \subsubsection{Transmit Rate and Latency}
    Since transmit rate and latency are critical performance metrics that can directly affect the overall quality and reliability of heterogeneous network communication, we mainly focus on optimizing them in this work. Based on all types of SINRs above, the transmit rates of different links over the same $h$-th sub-band can be hence established as \cite{shannon1948mathematical}, 
    \begin{equation}
        \begin{aligned}
            R^a[h] = W_b\log(1+\gamma^a[h]),
        \end{aligned}
    \end{equation}
    where $a\in\{LOW,HIGH\}$ represents different types of links and $b\in\{1,2\}$ is the bandwidth occupied for transmission, where $W_1$ and $W_2$ denote the bandwidth for low- and high-altitude links.
    
    
    
    \section{Optimization Objective}
    \label{sec:objective}
    In this section, we introduce two optimization models to assess how the SAGIN would operate, in which the ASO model is utilized to optimize over a range of possible latency probability distributions.
    ASO refers to the optimization of a system or process under conditions of uncertainty, where the probabilities of different outcomes or events are not precisely known~\cite{postek2018robust}.
    
    Given an object-power-channel decision profile $(\textbf{x},\textbf{p},\textbf{h})$ of this heterogeneous communication system, the ambiguous transmission rate $R: \mathcal{X} \times \mathcal{P} \times \mathcal{H} \times \Xi \rightarrow \mathbb{R}$, and a set of plausible distributions $\Omega$ over the uncertain parameters, the goal is to find the optimal decision profile $(\textbf{x}^*,\textbf{p}^*,\textbf{h}^*)$ that maximizes the worst-case expected transmission rate of $R$ over the set of plausible distributions,
    \begin{equation}
        \begin{aligned}
        &~~~~~~~\sup_{\mathbf{x},\mathbf{p},\mathbf{h}} \inf_{M \in \Omega} \mathbb{E}_{M} \bigg[  R^a(\mathbf{x},\mathbf{p},\mathbf{h}; \bm{\xi}) \bigg] \\
            \textrm{s.t.}~~~~& (a)~\sum_j x_{i,j}+x_{i,B}+x_{i,S}=1,\forall~GU_i,i\in\mathcal{N}_1,\\
            & (b)~x_{j,B}+x_{j,S}=1, \forall~UAV_j,j\in\mathcal{N}_2,\\
            & (c)~x_{i,j}, x_{i,B}, x_{i,S}, x_{j,B}, x_{j,S}\in \{0,1\}, \\
            & (d)~p_{i,j}, p_{i,B}, p_{i,S}, p_{j,B}, p_{j,S} \geq 0, 
        \end{aligned}
    \end{equation}
    where $\bm{\xi}$ some random variables representing the uncertain parameters, and $\mathbb{E}_M$ denotes the expected value with respect to the distribution $M$ such that $i\in\mathcal{N}_1$ and $j\in\mathcal{N}_2$. This formulation seeks to find a transmit decision that minimizes the maximum expected value of the objective function over all plausible distributions, accounting for the ambiguity and uncertainty in this SAGIN system.

    \section{Markov Decision Process of SAGIN}
    Considering the aforementioned SAGIN optimization objectives are non-convex functions and difficult to figure out with conventional programming methods, we model it as a Markov decision process (MDP), given as $\langle \boldsymbol{\mathcal{S}}, \boldsymbol{\mathcal{A}}, P, r, \beta\rangle$ for embedding the framework of MARL method, where $\boldsymbol{\mathcal{S}}\triangleq\{\mathcal{S}_1,\mathcal{S}_2\}$ and $\boldsymbol{\mathcal{A}}\triangleq\{\mathcal{A}_1,\mathcal{A}_2\}$ are the joint observation and action spaces of GU and UAV agent, respectively. $P$ indicates the transition probability, $r$ refers to the reward given by the SAGIN environment and $\beta$ represents the factor that determines the importance of future rewards.
    
    \subsection{Observation Space}
    In this SAGIN, we model GUs and UAVs as two types of agents that cooperatively explore the uncertain and dynamic communication environment. 
    \begin{figure}[t]
        \centering
        \includegraphics[width=\linewidth]{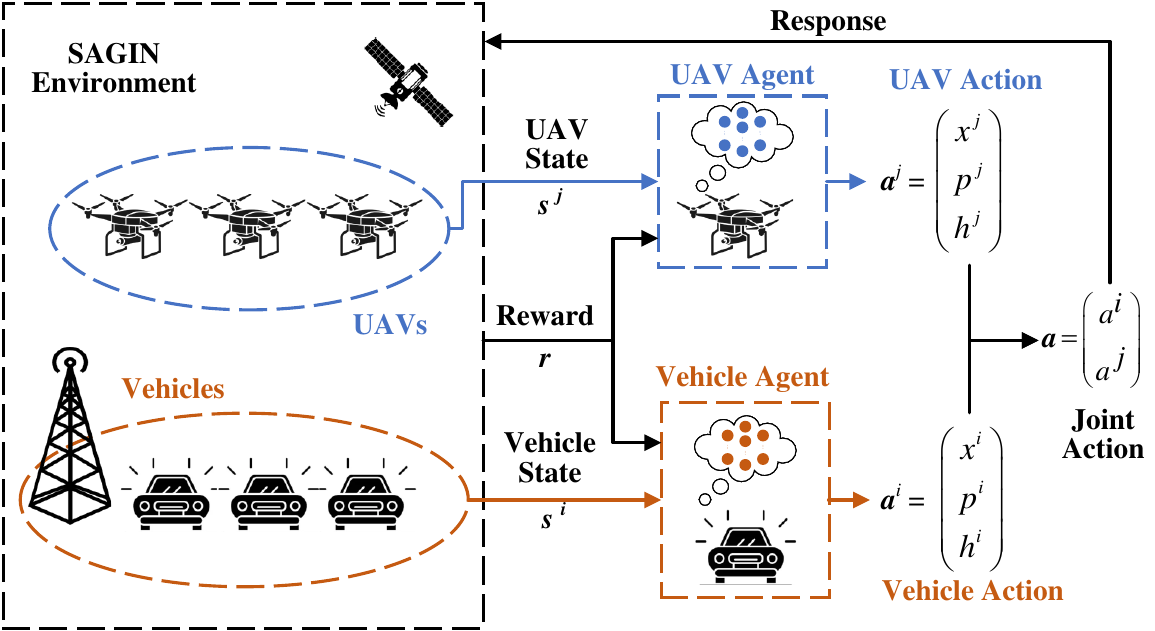}
        \caption{The workflow of Markov decision process in SAGIN.}
        \label{fig:reward}
    \end{figure}
    As discussed in Section~\ref{subsec:links}, SINR, on the one hand, is a measure of the quality of the received signal relative to the interference and noise. Agents can use the SINR to estimate the power of the received signal relative to the interference and noise, which can be used to make decisions about scheduling data transmissions. 
    On the other hand, interference power is a measure of the power of the interfering signal caused by other data transmissions. This metric can be utilized by agents to estimate the impact of other data transmissions on the quality of their own data transmissions.
    In addition, the data packet size is the size of the data packet that needs to be transmitted. Agents can leverage this information to estimate the time required to transmit the data packet and make decisions about scheduling data transmissions.
    Based on the above information, the observation space of agents can be defined as a vector of features that includes SINR $\gamma$, interference power $\phi$, and data packet size $B$. We hence establish the observation of GU agent $i$ as,
        $s^i = Y^i_{GU}(\boldsymbol{\gamma},\boldsymbol{\phi},\boldsymbol{B}) \in \mathcal{S}_1, \forall~GU_i,i\in\mathcal{N}_1,$
    nd similarly, the observation of UAV agent $j$ can be expressed as,
        $s^j = Y^j_{UAV}(\boldsymbol{\gamma},\boldsymbol{\phi},\boldsymbol{B}) \in \mathcal{S}_2,\forall~UAV_j,j\in\mathcal{N}_2$,
    where $Y_{GU}(\cdot)$ and $Y_{UAV}(\cdot)$ represent the observation functions of GU and UAV agents, respectively. In addition, the SINR $\boldsymbol{\gamma} = \{\boldsymbol{\gamma}^{LOW},\boldsymbol{\gamma}^{HIGH}\}$ and the interference power $\boldsymbol{\phi} = \{\boldsymbol{\phi}^{LOW},\boldsymbol{\phi}^{HIGH}\}$, both of which are modeled in Section~\ref{subsec:links}.

    \subsection{Action Space}
    The action space represents the set of feasible actions that can be taken by the GU or UAV agent in the SAGIN environment. The goal of each individual GU or UAV is to develop a collaborative policy that optimizes their collective long-term reward by selecting actions from their own designated action space.
    For each GU agent, it needs to select a receiver to transmit the data package at first, i.e. the communication link $x^i \in \mathcal{X}_1$, and then the appropriate communication channel $h^i \in \mathcal{H}_1$, and transmit power $p^i \in \mathcal{P}_1$ are supposed to be chosen for acceptable transmission efficiency.
    Therefore, each GU agent makes an object-power-channel decision profile,
        $a^i = (x^i,p^i,h^i) \in (\mathcal{X}_1,\mathcal{P}_1,\mathcal{H}_1) \subseteq \mathcal{A}_1,\forall~GU_i$,
    Similarly, the decision profile that each UAV agent at each time step is expressed as,
        $a^j = (x^j,p^j,h^j) \in (\mathcal{X}_2,\mathcal{P}_2,\mathcal{H}_2) \subseteq \mathcal{A}_2,\forall~UAV_j$.
    
    \subsection{Scenario Reward}
    Since the objective is to maximize the transmission rate of the SAGIN, we can naturally utilize the rate to design the reward that encourages agents to take actions that maximize the overall rate while still ensuring the stability and robustness of the network.
    Then, we stimulate coordination between agents. The reward should encourage agents to coordinate their actions to maximize the rate. 
    Third, the reward encourages exploration of the environment to discover new strategies that improve the rate. For example, agents can be rewarded for taking actions that have not been tried before, or for taking actions that improve the rate in a novel way.
    Notably, the robustness and performance need to be balanced.

    Based on these considerations, we specifically design the reward function for a cooperating MARL architecture in the aforementioned SAGIN as,
    \begin{equation}
        r = \alpha_1\cdot r_L + \alpha_2\cdot r_{coo} + \alpha_3\cdot r_{exp},
        \label{eq:reward}
    \end{equation}
    where $r$ is the common reward that all agents share, known as a cooperative reward. $\{\alpha_n\}_{n=1}^4$ are constant hyper-parameters that balance different reward components. $r_L$ stands for the latency of the network. To incentivize agents to minimize the latency of the SAGIN, we choose to use the negative of the latency as the reward. The higher the negative value of the latency, the greater the reward,
    \begin{equation}
        r_L = -(\kappa_1\sum_{i\in\mathcal{N}_1}L_i + \kappa_2\sum_{j\in\mathcal{N}_2}L_j),
        \label{eq:reward_latency}
    \end{equation}
    which ensures that agents will prioritize reducing the latency of the network. $\kappa_1$ and $\kappa_1$ are two hyper-parameters that balance the latencies of different communication devices.
    And $r_{coo}$ is a reward for actions that improve coordination between agents. For encouraging collaboration and coordination between Gus and UAVs to minimize latency, we give a reward to agents who take actions contributing to a coordinated strategy that reduces latency,
    \begin{equation}
        r_{coo} = \sum_{i\in\mathcal{N}_1} w_i\delta_i + \sum_{j\in\mathcal{N}_2}w_j \delta_j,
        \label{eq:reward_coordination}
    \end{equation}
    where $w_i$ and $w_j$ refer to the weights assigned to GU $i$ and UAV $j$ indicating the importance of each agent's role in the network. $\delta_i$ and $\delta_j$ are binary indicators of whether their actions contributed to a coordinated strategy that reduced latency. To measure the reduction in latency resulting from a coordinated strategy, we utilize a metric such as the difference between the latency before and after the strategy is implemented. If the coordinated strategy results in a lower latency compared to the latency without the strategy, then we can consider that the action of GU $i$ or UAV $j$ contributes to a coordinated strategy that reduces the average latency of the SAGIN system $\Bar{L}$, designed as,
    \begin{equation}
        \delta = 
        \begin{cases}
          1, & \text{if $\Bar{L}_t$ $-$ $\Bar{L}_{t+1}$ $>$ 0},\\
          0, & \text{otherwise}.
        \end{cases}
        \label{eq:delta}
    \end{equation}
    Further, $r_{exp}$ indicates a reward for novel actions that improve latency. To motivate agents to explore the environment and try new strategies, we can use an exploration bonus that rewards agents for taking actions that have not been tried before or that improve latency in a novel way. One way to do this is to give a reward to agents that take actions that increase the diversity of strategies used to reduce latency,
    \begin{equation}
        r_{exp} = \zeta_1 \sum_{i\in\mathcal{N}_1}\mathcal{H}^i(\pi^i)+\zeta_2 \sum_{j\in\mathcal{N}_2}\mathcal{H}^j(\pi^j),
        \label{eq:reward_exploration}
    \end{equation}
    where $\zeta_1$ and $\zeta_2$ are two weighting factors that control the influence of the exploration bonus, and $\mathcal{H}(\pi)$ is the entropy of the probability distribution $\pi$ over the set of strategies used to reduce latency. We in this article utilize the Shannon entropy as a measure of the uncertainty or randomness of a probability distribution, given as $\mathcal{H}(\pi) = - \sum_{a\in\mathcal{A}} \pi(a) log_2 \pi(a)$, in which $\pi(a)$ is the probability of selecting action $a$ over its action space $\mathcal{A}$. This encourages agents to explore different strategies and avoid relying on a single strategy that may not be effective in all situations.

    \section{Experiments and Results}
    \begin{table}[t]
		\begin{center}
			\caption{Simulation settings \cite{winnerII,3gpp36777,6863654,3gpp38811}}
			\setlength{\tabcolsep}{0mm}{
				\begin{tabular}{cc}
					\toprule  
					\textbf{SAGIN parameter} & \textbf{Value} \\
					\midrule  
					Number of vehicle & $\{2,5,10,20,30\}$ \\
					Number of UAV & $\{1,2,5,10,15\}$ \\
					Number of BS & 1 \\
					Number of satellite & 1 \\
                    Environment area & 200$\times$100 m$^2$ \\
                    Height of vehicle & 1.5 m\\
                    Height of UAV & 25 m\\
                    Height of BS & 100 m\\
                    Height of satellite & 600 km\\
                    Angle of satellite & 30$\degree$\\
					Carrier frequency of terrestrial links & 2 GHz \\
					Bandwidth  of terrestrial links & 1 MHz \\
					Carrier frequency of non-terrestrial links & 30 GHz \\
					Bandwidth of non-terrestrial links & 100 MHz \\
					GU \& UAV transmit power & $\{23,10,5,-100\}$ dBm \\
					BS antenna gain & 8 dBi\\
					UAV \& GU antenna gain & 3 dBi\\
					GU \& UAV receiver noise figure & 9 dB\\
					BS receiver noise figure & 5 dB\\
					Noise power & -114 dBm\\
					Time slot for package delivery & 100 ms \\
                \bottomrule
				\end{tabular}
				\label{tab:sim_setting}}
				\vspace{-.3cm}
		\end{center}
	\end{table}
    For implementing the proposed CMT-MARL method in the SAGIN environment, we have novelly designed a SAGIN communication scenario, in which different channel models to characterize different channels in SAGIN are combined. Elaborately, the G2B channel is modeled using the WINNER II channel model~\cite{winnerII}, the U2B channel is modeled according to the definition provided in 3GPP TR 36.777 Rel. 15~\cite{3gpp36777}, the G2U channel follows the definition in \cite{6863654}, and the non-terrestrial links (G2S and U2S) are modeled based on the definition in 3GPP TR 38.811 Rel. 15~\cite{3gpp38811}. The specific simulation experiment settings can be found in Table~\ref{tab:sim_setting}.

    To evaluate the performance of the proposed CMT-MARL method in the training phase, we first employ 2 vehicles and 1 UAV and exhibit the accumulated reward of the SAGIN system, in which the reward is normalized within the range of [0, 1] to effectively showcase the capabilities of the multi-type multi-agent system. Each episode comprises 100 steps, corresponding to a duration of 100 ms, during which the vehicle and UAV agents transmit packages. The convergence of system performance can be observed in Figure~\ref{fig:reward}, where it becomes apparent that the performance stabilizes after approximately 500 episodes. The baseline represents that the vehicle and UAV agents randomly select actions at any time.
    \begin{figure}[t]
        \centering
        \includegraphics[width=.8\linewidth]{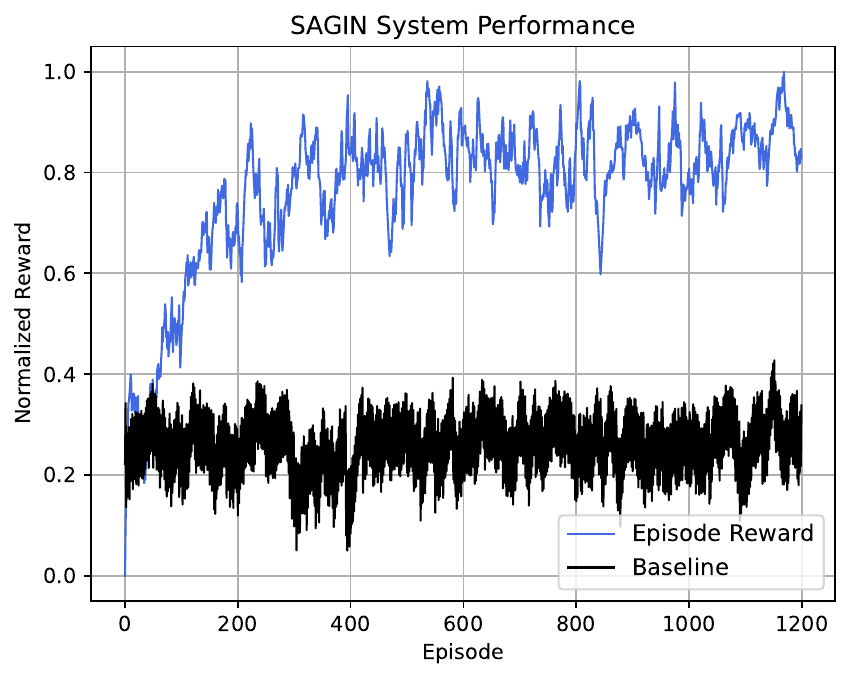}
        \caption{SAGIN system performance.}
        \label{fig:reward}
        \vspace{-.4cm}
    \end{figure}
    
    \begin{figure}[t]
        \centering
        \includegraphics[width=.8\linewidth]{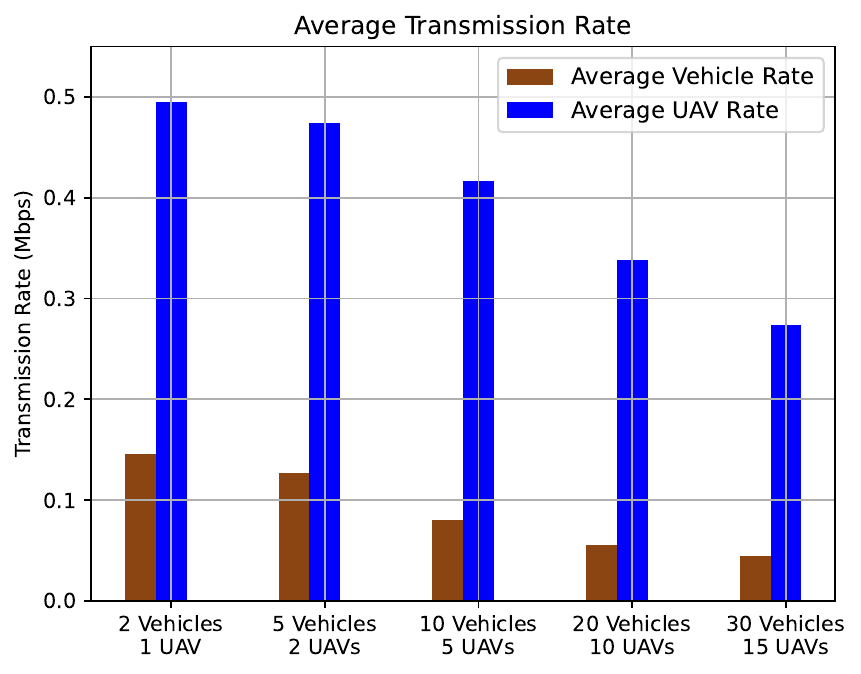}
        \caption{Average transmission rates with different numbers of vehicles and UAVs.}
        \label{fig:average transmission rates}
        \vspace{-.4cm}
    \end{figure}

    \begin{figure}[t]
        \centering
        \includegraphics[width=.8\linewidth]{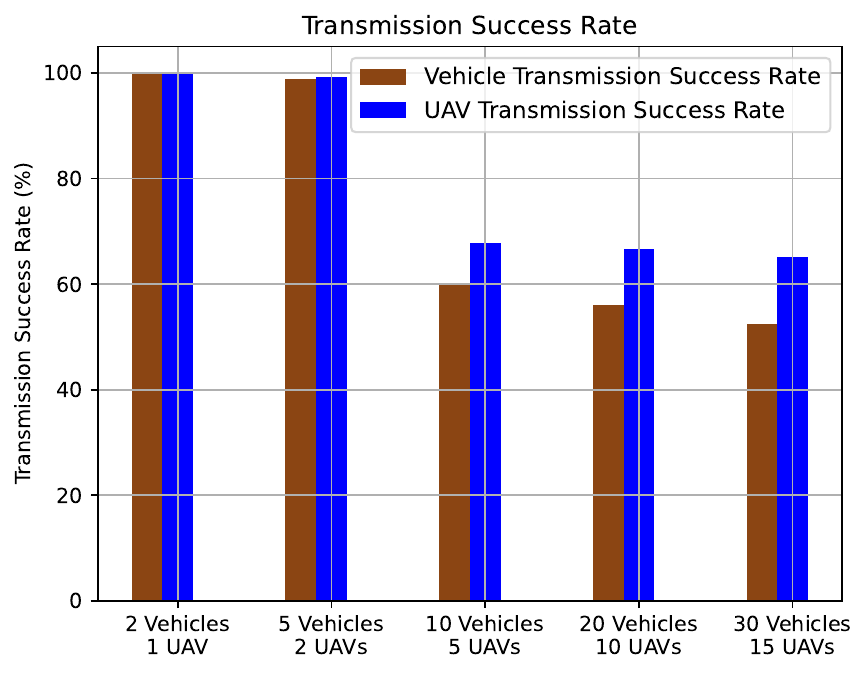}
        \caption{Transmission success rates with different numbers of vehicles and UAVs.}
        \label{fig:success rates}
        \vspace{-.4cm}
    \end{figure}

    During the test phase, we assess the effectiveness of the CMT-MARL approach by varying the quantities of vehicle-UAV combinations, while maintaining a constant number of one base station (BS) and one satellite throughout the evaluations, as listed in Table~\ref{tab:sim_setting}. The test area spans 200$\times$100 m$^2$. Figure~\ref{fig:average transmission rates} illustrates the average transmission rates achieved by the vehicle and UAV groups. The results show that the vehicle group attains an average transmission rate of approximately 0.15 Mbps, while the UAV group achieves around 0.50 Mbps. Notably, even when the number of agents increases to 30 vehicles and 15 UAVs, indicating a crowded communication environment, the average transmission rates remain relatively stable at approximately 0.05 Mbps for the vehicle group and 0.28 Mbps for the UAV group.
     
    Furthermore, in Figure~\ref{fig:success rates}, we demonstrate the transmission efficacy as reflected by the success rates across varying quantities of vehicle-UAV combinations. As the agent amount rises, the escalating interference within the SAGIN system poses hurdles to the transmissions. Consequently, agents must endeavor to discern and implement more efficient transmission policies to overcome the intensified interference and successfully accomplish their package transmission tasks. Hereto, The CMT-MARL technique allows the SAGIN system to achieve a nearly 100\% success rate with a relatively smaller number of agents, i.e., $\{2,1\}$ and $\{5,2\}$ vehicle-UAV combinations. As the agent number expands from $\{10,5\}$ to $\{30,15\}$, the success rates for both vehicle and UAV agents remain consistently high, hovering around 55\% to 60\% for the former and approximately 65\% for the latter. This attests to the robustness and effectiveness of the CMT-MARL technique in facilitating reliable and successful transmissions within the SAGIN environment.

    \section{Conclusion}
    In this study, to elaborately investigate the resource management problem within the SAGIN system, we have innovatively constructed a SAGIN system incorporating five distinct communication links and proposed a specific MARL technique called CMT-MARL to solve it. This method is tested using different amounts of vehicle-UAV agent combinations and showcases its reliability and robustness in the SAGIN, providing the hope of implementing it over the physical SAGIN environment. In future works, distributed decision-making, as well as the scalability and efficiency issues, will further be investigated using mean field and MARL methods in consideration that each vehicle or UAV agent can make decisions based on local observations and interactions with their immediate environment, without requiring centralized control. This distributed approach allows the network to scale effectively as the number of agents increases.

\begin{acks}
This work was supported in part by Shenzhen Science and Technology Program (JCYJ20220530143013030), and by Tsinghua Shenzhen International Graduate School-Shenzhen Pengrui Young Faculty Program of Shenzhen Pengrui Foundation (No. SZPR2023005).
\end{acks}
\bibliographystyle{ACM-Reference-Format}
\bibliography{refs_cpd.bib}


\begin{thebibliography}{27}


\ifx \showCODEN    \undefined \def \showCODEN     #1{\unskip}     \fi
\ifx \showDOI      \undefined \def \showDOI       #1{#1}\fi
\ifx \showISBNx    \undefined \def \showISBNx     #1{\unskip}     \fi
\ifx \showISBNxiii \undefined \def \showISBNxiii  #1{\unskip}     \fi
\ifx \showISSN     \undefined \def \showISSN      #1{\unskip}     \fi
\ifx \showLCCN     \undefined \def \showLCCN      #1{\unskip}     \fi
\ifx \shownote     \undefined \def \shownote      #1{#1}          \fi
\ifx \showarticletitle \undefined \def \showarticletitle #1{#1}   \fi
\ifx \showURL      \undefined \def \showURL       {\relax}        \fi
\providecommand\bibfield[2]{#2}
\providecommand\bibinfo[2]{#2}
\providecommand\natexlab[1]{#1}
\providecommand\showeprint[2][]{arXiv:#2}

\bibitem[3gp(2017)]%
        {3gpp36777}
 \bibinfo{year}{2017}\natexlab{}.
\newblock \bibinfo{journal}{\emph{3rd Generation Partnership Project; Technical
  Specification Group Radio Access Network;Study on Enhanced LTE Support for
  Aerial Vehicles: (Release 15)}} (\bibinfo{date}{Standard 3GPP TR, Dec}
  \bibinfo{year}{2017}).
\newblock


\bibitem[3gp(2020)]%
        {3gpp38811}
 \bibinfo{year}{2020}\natexlab{}.
\newblock \bibinfo{journal}{\emph{3rd Generation Partnership Project; Technical
  Specification Group Radio Access Network; Study on New Radio (NR) to support
  non-terrestrial networks : (Release 15)}} (\bibinfo{date}{Standard 3GPP TR,
  Sep} \bibinfo{year}{2020}).
\newblock


\bibitem[Agiwal et~al\mbox{.}(2016)]%
        {agiwal2016next}
\bibfield{author}{\bibinfo{person}{Mamta Agiwal}, \bibinfo{person}{Abhishek
  Roy}, {and} \bibinfo{person}{Navrati Saxena}.}
  \bibinfo{year}{2016}\natexlab{}.
\newblock \showarticletitle{Next generation 5G wireless networks: A
  comprehensive survey}.
\newblock \bibinfo{journal}{\emph{IEEE Communications Surveys \& Tutorials}}
  \bibinfo{volume}{18}, \bibinfo{number}{3} (\bibinfo{year}{2016}),
  \bibinfo{pages}{1617--1655}.
\newblock


\bibitem[Al-Fuqaha et~al\mbox{.}(2015)]%
        {IoT}
\bibfield{author}{\bibinfo{person}{Ala Al-Fuqaha}, \bibinfo{person}{Mohsen
  Guizani}, \bibinfo{person}{Mehdi Mohammadi}, \bibinfo{person}{Mohammed
  Aledhari}, {and} \bibinfo{person}{Moussa Ayyash}.}
  \bibinfo{year}{2015}\natexlab{}.
\newblock \showarticletitle{Internet of Things: A Survey on Enabling
  Technologies, Protocols, and Applications}.
\newblock \bibinfo{journal}{\emph{IEEE Communications Surveys \& Tutorials}}
  \bibinfo{volume}{17}, \bibinfo{number}{4} (\bibinfo{year}{2015}),
  \bibinfo{pages}{2347--2376}.
\newblock


\bibitem[Al-Hourani et~al\mbox{.}(2014)]%
        {6863654}
\bibfield{author}{\bibinfo{person}{Akram Al-Hourani},
  \bibinfo{person}{Sithamparanathan Kandeepan}, {and} \bibinfo{person}{Simon
  Lardner}.} \bibinfo{year}{2014}\natexlab{}.
\newblock \showarticletitle{Optimal LAP Altitude for Maximum Coverage}.
\newblock \bibinfo{journal}{\emph{IEEE Wireless Communications Letters}}
  \bibinfo{volume}{3}, \bibinfo{number}{6} (\bibinfo{year}{2014}),
  \bibinfo{pages}{569--572}.
\newblock


\bibitem[Albuquerque et~al\mbox{.}(2007)]%
        {albuquerque2007global}
\bibfield{author}{\bibinfo{person}{Marcelo Albuquerque}, \bibinfo{person}{Arun
  Ayyagari}, \bibinfo{person}{Michael~A Dorsett}, {and}
  \bibinfo{person}{Michael~S Foster}.} \bibinfo{year}{2007}\natexlab{}.
\newblock \showarticletitle{Global information grid (GIG) edge network
  interface architecture}. In \bibinfo{booktitle}{\emph{MILCOM 2007-IEEE
  Military Communications Conference}}. IEEE, \bibinfo{pages}{1--7}.
\newblock


\bibitem[Andrews et~al\mbox{.}(2014)]%
        {andrews2014will}
\bibfield{author}{\bibinfo{person}{Jeffrey~G Andrews}, \bibinfo{person}{Stefano
  Buzzi}, \bibinfo{person}{Wan Choi}, \bibinfo{person}{Stephen~V Hanly},
  \bibinfo{person}{Angel Lozano}, \bibinfo{person}{Anthony~CK Soong}, {and}
  \bibinfo{person}{Jianzhong~Charlie Zhang}.} \bibinfo{year}{2014}\natexlab{}.
\newblock \showarticletitle{What will 5G be?}
\newblock \bibinfo{journal}{\emph{IEEE Journal on selected areas in
  communications}} \bibinfo{volume}{32}, \bibinfo{number}{6}
  (\bibinfo{year}{2014}), \bibinfo{pages}{1065--1082}.
\newblock


\bibitem[Casoni et~al\mbox{.}(2015)]%
        {casoni2015integration}
\bibfield{author}{\bibinfo{person}{Maurizio Casoni},
  \bibinfo{person}{Carlo~Augusto Grazia}, \bibinfo{person}{Martin Klapez},
  \bibinfo{person}{Natale Patriciello}, \bibinfo{person}{Angelos Amditis},
  {and} \bibinfo{person}{Evangelos Sdongos}.} \bibinfo{year}{2015}\natexlab{}.
\newblock \showarticletitle{Integration of satellite and LTE for disaster
  recovery}.
\newblock \bibinfo{journal}{\emph{IEEE Communications Magazine}}
  \bibinfo{volume}{53}, \bibinfo{number}{3} (\bibinfo{year}{2015}),
  \bibinfo{pages}{47--53}.
\newblock


\bibitem[Chandrasekharan et~al\mbox{.}(2016)]%
        {chandrasekharan2016designing}
\bibfield{author}{\bibinfo{person}{Sathyanarayanan Chandrasekharan},
  \bibinfo{person}{Karina Gomez}, \bibinfo{person}{Akram Al-Hourani},
  \bibinfo{person}{Sithamparanathan Kandeepan}, \bibinfo{person}{Tinku
  Rasheed}, \bibinfo{person}{Leonardo Goratti}, \bibinfo{person}{Laurent
  Reynaud}, \bibinfo{person}{David Grace}, \bibinfo{person}{Isabelle Bucaille},
  \bibinfo{person}{Thomas Wirth}, {et~al\mbox{.}}}
  \bibinfo{year}{2016}\natexlab{}.
\newblock \showarticletitle{Designing and implementing future aerial
  communication networks}.
\newblock \bibinfo{journal}{\emph{IEEE Communications Magazine}}
  \bibinfo{volume}{54}, \bibinfo{number}{5} (\bibinfo{year}{2016}),
  \bibinfo{pages}{26--34}.
\newblock


\bibitem[Döttling et~al\mbox{.}(2010)]%
        {winnerII}
\bibfield{author}{\bibinfo{person}{Martin Döttling}, \bibinfo{person}{Werner
  Mohr}, {and} \bibinfo{person}{Afif Osseiran}.}
  \bibinfo{year}{2010}\natexlab{}.
\newblock \bibinfo{booktitle}{\emph{WINNER II Channel Models}}.
\newblock \bibinfo{pages}{39--92}.
\newblock


\bibitem[El~Tanab and Hamouda(2016)]%
        {el2016resource}
\bibfield{author}{\bibinfo{person}{Manal El~Tanab} {and} \bibinfo{person}{Walaa
  Hamouda}.} \bibinfo{year}{2016}\natexlab{}.
\newblock \showarticletitle{Resource allocation for underlay cognitive radio
  networks: A survey}.
\newblock \bibinfo{journal}{\emph{IEEE Communications Surveys \& Tutorials}}
  \bibinfo{volume}{19}, \bibinfo{number}{2} (\bibinfo{year}{2016}),
  \bibinfo{pages}{1249--1276}.
\newblock


\bibitem[Farserotu and Prasad(2000)]%
        {farserotu2000survey}
\bibfield{author}{\bibinfo{person}{John Farserotu} {and}
  \bibinfo{person}{Ramjee Prasad}.} \bibinfo{year}{2000}\natexlab{}.
\newblock \showarticletitle{A survey of future broadband multimedia satellite
  systems, issues and trends}.
\newblock \bibinfo{journal}{\emph{IEEE Communications Magazine}}
  \bibinfo{volume}{38}, \bibinfo{number}{6} (\bibinfo{year}{2000}),
  \bibinfo{pages}{128--133}.
\newblock


\bibitem[Foerster et~al\mbox{.}(2016)]%
        {foerster2016learning}
\bibfield{author}{\bibinfo{person}{Jakob Foerster},
  \bibinfo{person}{Ioannis~Alexandros Assael}, \bibinfo{person}{Nando
  De~Freitas}, {and} \bibinfo{person}{Shimon Whiteson}.}
  \bibinfo{year}{2016}\natexlab{}.
\newblock \showarticletitle{Learning to communicate with deep multi-agent
  reinforcement learning}.
\newblock \bibinfo{journal}{\emph{Advances in neural information processing
  systems}}  \bibinfo{volume}{29} (\bibinfo{year}{2016}).
\newblock


\bibitem[Hubenko et~al\mbox{.}(2006)]%
        {hubenko2006improving}
\bibfield{author}{\bibinfo{person}{Victor~P Hubenko},
  \bibinfo{person}{Richard~A Raines}, \bibinfo{person}{Robert~F Mills},
  \bibinfo{person}{Rusty~O Baldwin}, \bibinfo{person}{Barry~E Mullins}, {and}
  \bibinfo{person}{Michael~R Grimaila}.} \bibinfo{year}{2006}\natexlab{}.
\newblock \showarticletitle{Improving the global information grid's performance
  through satellite communications layer enhancements}.
\newblock \bibinfo{journal}{\emph{IEEE Communications Magazine}}
  \bibinfo{volume}{44}, \bibinfo{number}{11} (\bibinfo{year}{2006}),
  \bibinfo{pages}{66--72}.
\newblock


\bibitem[Kato et~al\mbox{.}(2019)]%
        {kato2019optimizing}
\bibfield{author}{\bibinfo{person}{Nei Kato}, \bibinfo{person}{Zubair~Md
  Fadlullah}, \bibinfo{person}{Fengxiao Tang}, \bibinfo{person}{Bomin Mao},
  \bibinfo{person}{Shigenori Tani}, \bibinfo{person}{Atsushi Okamura}, {and}
  \bibinfo{person}{Jiajia Liu}.} \bibinfo{year}{2019}\natexlab{}.
\newblock \showarticletitle{Optimizing space-air-ground integrated networks by
  artificial intelligence}.
\newblock \bibinfo{journal}{\emph{IEEE Wireless Communications}}
  \bibinfo{volume}{26}, \bibinfo{number}{4} (\bibinfo{year}{2019}),
  \bibinfo{pages}{140--147}.
\newblock


\bibitem[Liang et~al\mbox{.}(2019)]%
        {liang2019spectrum}
\bibfield{author}{\bibinfo{person}{Le Liang}, \bibinfo{person}{Hao Ye}, {and}
  \bibinfo{person}{Geoffrey~Ye Li}.} \bibinfo{year}{2019}\natexlab{}.
\newblock \showarticletitle{Spectrum sharing in vehicular networks based on
  multi-agent reinforcement learning}.
\newblock \bibinfo{journal}{\emph{IEEE Journal on Selected Areas in
  Communications}} \bibinfo{volume}{37}, \bibinfo{number}{10}
  (\bibinfo{year}{2019}), \bibinfo{pages}{2282--2292}.
\newblock


\bibitem[Liu et~al\mbox{.}(2018)]%
        {liu2018space}
\bibfield{author}{\bibinfo{person}{Jiajia Liu}, \bibinfo{person}{Yongpeng Shi},
  \bibinfo{person}{Zubair~Md Fadlullah}, {and} \bibinfo{person}{Nei Kato}.}
  \bibinfo{year}{2018}\natexlab{}.
\newblock \showarticletitle{Space-air-ground integrated network: A survey}.
\newblock \bibinfo{journal}{\emph{IEEE Communications Surveys \& Tutorials}}
  \bibinfo{volume}{20}, \bibinfo{number}{4} (\bibinfo{year}{2018}),
  \bibinfo{pages}{2714--2741}.
\newblock


\bibitem[Postek et~al\mbox{.}(2018)]%
        {postek2018robust}
\bibfield{author}{\bibinfo{person}{Krzysztof Postek}, \bibinfo{person}{Aharon
  Ben-Tal}, \bibinfo{person}{Dick Den~Hertog}, {and} \bibinfo{person}{Bertrand
  Melenberg}.} \bibinfo{year}{2018}\natexlab{}.
\newblock \showarticletitle{Robust optimization with ambiguous stochastic
  constraints under mean and dispersion information}.
\newblock \bibinfo{journal}{\emph{Operations Research}} \bibinfo{volume}{66},
  \bibinfo{number}{3} (\bibinfo{year}{2018}), \bibinfo{pages}{814--833}.
\newblock


\bibitem[Pultarova(2015)]%
        {pultarova2015telecommunications}
\bibfield{author}{\bibinfo{person}{Tereza Pultarova}.}
  \bibinfo{year}{2015}\natexlab{}.
\newblock \showarticletitle{Telecommunications-space tycoons go head to head
  over mega satellite network [news briefing]}.
\newblock \bibinfo{journal}{\emph{Engineering \& Technology}}
  \bibinfo{volume}{10}, \bibinfo{number}{2} (\bibinfo{year}{2015}),
  \bibinfo{pages}{20--20}.
\newblock


\bibitem[Shang et~al\mbox{.}(2021)]%
        {shang2021computing}
\bibfield{author}{\bibinfo{person}{Bodong Shang}, \bibinfo{person}{Yang Yi},
  {and} \bibinfo{person}{Lingjia Liu}.} \bibinfo{year}{2021}\natexlab{}.
\newblock \showarticletitle{Computing over space-air-ground integrated
  networks: Challenges and opportunities}.
\newblock \bibinfo{journal}{\emph{IEEE Network}} \bibinfo{volume}{35},
  \bibinfo{number}{4} (\bibinfo{year}{2021}), \bibinfo{pages}{302--309}.
\newblock


\bibitem[Shannon(1948)]%
        {shannon1948mathematical}
\bibfield{author}{\bibinfo{person}{Claude~E Shannon}.}
  \bibinfo{year}{1948}\natexlab{}.
\newblock \showarticletitle{A mathematical theory of communication}.
\newblock \bibinfo{journal}{\emph{The Bell system technical journal}}
  \bibinfo{volume}{27}, \bibinfo{number}{3} (\bibinfo{year}{1948}),
  \bibinfo{pages}{379--423}.
\newblock


\bibitem[Subramanian et~al\mbox{.}(2020)]%
        {subramanian2020multi}
\bibfield{author}{\bibinfo{person}{Sriram~Ganapathi Subramanian},
  \bibinfo{person}{Pascal Poupart}, \bibinfo{person}{Matthew~E Taylor}, {and}
  \bibinfo{person}{Nidhi Hegde}.} \bibinfo{year}{2020}\natexlab{}.
\newblock \showarticletitle{Multi type mean field reinforcement learning}.
\newblock \bibinfo{journal}{\emph{arXiv preprint arXiv:2002.02513}}
  (\bibinfo{year}{2020}).
\newblock


\bibitem[Sutton and Barto(2018)]%
        {sutton2018reinforcement}
\bibfield{author}{\bibinfo{person}{Richard~S Sutton} {and}
  \bibinfo{person}{Andrew~G Barto}.} \bibinfo{year}{2018}\natexlab{}.
\newblock \bibinfo{booktitle}{\emph{Reinforcement learning: An introduction}}.
\newblock \bibinfo{publisher}{MIT press}.
\newblock


\bibitem[Xiong et~al\mbox{.}(2016)]%
        {xiong2016kind}
\bibfield{author}{\bibinfo{person}{Gang Xiong}, \bibinfo{person}{Fenghua Zhu},
  \bibinfo{person}{Xisong Dong}, \bibinfo{person}{Haisheng Fan},
  \bibinfo{person}{Bin Hu}, \bibinfo{person}{Qingjie Kong},
  \bibinfo{person}{Wenwen Kang}, {and} \bibinfo{person}{Teng Teng}.}
  \bibinfo{year}{2016}\natexlab{}.
\newblock \showarticletitle{A kind of novel ITS based on space-air-ground
  big-data}.
\newblock \bibinfo{journal}{\emph{IEEE intelligent transportation systems
  magazine}} \bibinfo{volume}{8}, \bibinfo{number}{1} (\bibinfo{year}{2016}),
  \bibinfo{pages}{10--22}.
\newblock


\bibitem[Zhang et~al\mbox{.}(2021)]%
        {zhang2021multi}
\bibfield{author}{\bibinfo{person}{Kaiqing Zhang}, \bibinfo{person}{Zhuoran
  Yang}, {and} \bibinfo{person}{Tamer Ba{\c{s}}ar}.}
  \bibinfo{year}{2021}\natexlab{}.
\newblock \showarticletitle{Multi-agent reinforcement learning: A selective
  overview of theories and algorithms}.
\newblock \bibinfo{journal}{\emph{Handbook of reinforcement learning and
  control}} (\bibinfo{year}{2021}), \bibinfo{pages}{321--384}.
\newblock


\bibitem[Zhang et~al\mbox{.}(2017)]%
        {zhang2017software}
\bibfield{author}{\bibinfo{person}{Ning Zhang}, \bibinfo{person}{Shan Zhang},
  \bibinfo{person}{Peng Yang}, \bibinfo{person}{Omar Alhussein},
  \bibinfo{person}{Weihua Zhuang}, {and} \bibinfo{person}{Xuemin~Sherman
  Shen}.} \bibinfo{year}{2017}\natexlab{}.
\newblock \showarticletitle{Software defined space-air-ground integrated
  vehicular networks: Challenges and solutions}.
\newblock \bibinfo{journal}{\emph{IEEE Communications Magazine}}
  \bibinfo{volume}{55}, \bibinfo{number}{7} (\bibinfo{year}{2017}),
  \bibinfo{pages}{101--109}.
\newblock


\bibitem[Zhou et~al\mbox{.}(2020)]%
        {zhou2020deep}
\bibfield{author}{\bibinfo{person}{Conghao Zhou}, \bibinfo{person}{Wen Wu},
  \bibinfo{person}{Hongli He}, \bibinfo{person}{Peng Yang},
  \bibinfo{person}{Feng Lyu}, \bibinfo{person}{Nan Cheng}, {and}
  \bibinfo{person}{Xuemin Shen}.} \bibinfo{year}{2020}\natexlab{}.
\newblock \showarticletitle{Deep reinforcement learning for delay-oriented IoT
  task scheduling in SAGIN}.
\newblock \bibinfo{journal}{\emph{IEEE Transactions on Wireless
  Communications}} \bibinfo{volume}{20}, \bibinfo{number}{2}
  (\bibinfo{year}{2020}), \bibinfo{pages}{911--925}.
\newblock


\end{thebibliography}
 
\end{document}